\documentclass[review]{elsarticle}
\usepackage{amssymb}

\newcommand{\hL}{\hat{\cal L}}
\newcommand{\hF}{\hat{\cal F}}

\newcommand{\prt}{\partial}

\newcommand{\rgl}{\rangle}
\newcommand{\lgl}{\langle}

\newcommand{\vep}{\varepsilon}

\newcommand{\be}{\begin{equation}}
\newcommand{\ee}{\end{equation}}
\newcommand{\bea}{\begin{eqnarray}}
\newcommand{\eea}{\end{eqnarray}}

\begin{document}

\begin{frontmatter}

\title{L\'evy flights in a box}

\author{Alexander Iomin }
\ead{iomin@physics.technion.ac.il}

%\cortext[cor]{Corresponding author}

\address{Department of Physics,
Technion, Haifa, 32000, Israel}

%\address{Max-Planck-Institute for Physics of Complex Systems,
%Dresden, Germany }

\begin{abstract}
It is shown that a quantum L\'evy process in a box leads to a
problem involving topological constraints in space, and its
treatment in the framework of the path integral formalism with the
L\'evy measure is suggested.  The eigenvalue problem for the
infinite potential well is properly defined and solved. An
analytical expression for the evolution operator is obtained in
the path integral presentation, and the path integral takes the
correct limit of the local quantum mechanics with topological
constraints. An example of the L\'evy process in oscillating walls
is also considered in the adiabatic approximation.
%
%as well, and an analytical expression for the quasienergy spectrum
%is obtained
%%%

%
\end{abstract}

\begin{keyword}
Fractional integral\sep L\'evy flight\sep Topological constraints
\sep Path integral


\end{keyword}

\end{frontmatter}

\section{Introduction}\label{sec:int}

The introduction of a fractional concept in quantum mechanics with
motivating new implementations of non-local physics leads to many
technical questions and often needs special care. A typical
example is the ``quantum L\'evy flights'' of a particle in an
infinite potential well, suggested in \cite{laskin1}. This
``simple'' example has evoked an active discussion in the
literature \cite{Jeng,Bayin1,Bayin2,Hawkins,luchko,herrmann} on
how a non-local operator, defined on the infinite scale, acts in a
finite-size range, such as a quantum box. The aim of the present
research, related to this question, is an exploration of the
L\'evy flights in boundary value problems in a finite-size area.
Among many possible applications of the problem, of special
interest is the first-passage analysis \cite{Dubiec}, which is
important for the investigation of the transport properties of
L\'evy glasses \cite{Barthelemy}.

In quantum mechanics, the fractional concept has been introduced
by means of the Feynman propagator for non-relativistic quantum
mechanics as for Brownian path integrals \cite{laskin1,west}.
Equivalence between the Wiener and the Feynman path integrals
\cite{feynman}, established by Kac \cite{kac}, is natural, since
both are Markov processes.
%indicates equivalence between the Laplace operators for classical diffusion
%equation and the Schr\"odinger equation.
%The same, the appearance of the space fractional derivatives in
%the Schr\"odinger equation is natural, since both the standard
%Schr\"odinger equation and the space fractional one obey the
%Markov process.
As shown in the seminal papers \cite{laskin1, west}, the
appearance of the space fractional derivatives in the
Schr\"odinger equation is natural and relates to the path
integrals approach. The path integral approach for L\'evy stable
processes, considered for the fractional diffusion equation, in
particular, for the fractional Fokker-Planck equation (FFPE),  has
been extended to a quantum Feynman-L\'evy measure that leads to
the space fractional Schr\"odinger equation (FSE)
\cite{laskin1,west}.

%Note also that the FFPE is well justified from the physical point
%of view as an asymptotic description of the L\'evy random
%processes, considered in the framework of the continuous time
%random walks (CTRW) (see reviews \cite{shlesinger,klafter}).

The introduction of the L\'evy measure in quantum mechanics is
based on the generalization of the self-consistency condition,
known as the Bachelor-Smoluchowski-Kol\-mo\-go\-rov chain equation
(or the Einstein-Smoluchowski-Kol\-mo\-go\-rov-Chapman equation,
see \textit{e.g.}, \cite{shlesinger}), established for the Wiener
process for the conditional probability $W(x,t|x_0,t_0)$
\be\label{Smoluch} %
W(x,t|x_0,t_0)=\int_{-\infty}^{\infty}W(x,t|x',t')W(x',t'|x_0,t_0)\,
dx'\,  . \ee%
In the case of the translational symmetry, it reads
$W(x,t|x_0,t_0)=W(x-x_0,t|t_0)$. Straightforward generalization of
this expression by the L\'evy process is expressed through the
Fourier transform
\be\label{levy1}  %
W(x,t|x_0,t_0)=\int_{-\infty}^{\infty}
e^{ip(x-x_0)}e^{-K_{\alpha}t|p|^{\alpha}}\,dp\, ,  \ee %
where $0<\alpha\leq 2$ and $K_{\alpha}$ is a generalized diffusion
coefficient \cite{klafter}. Using Eqs. (\ref{Smoluch}) and
(\ref{levy1}), one defines the integration of function
$F[x(\tau)]$ over the generalized measure
\bea\label{levy2}  %
\int F[x(\tau)]d_Lx(\tau)
\equiv\lim_{N\to\infty}\frac{1}{(2\pi\hbar)^{n}}
\int_{-\infty}^{\infty}dx_1\dots\int_{-\infty}^{\infty}dx_N
\int_{-\infty}^{\infty}dp_1\dots\int_{-\infty}^{\infty}dp_N
\nonumber \\
\times F(x_1,\dots\, , x_N)
e^{-i[(x_1-x_0)p_1+\dots+(x_{N+1}-x_N)p_N]}
e^{-K_{\alpha}[|p_1|^{\alpha}\Delta t +\dots +
|p_N|^{\alpha}\Delta t]}\, .  %
\eea  %
Here, $\Delta t=t/N$. The L\'evy distribution is defined by the
Fox function \cite{laskin1} and in the r.h.s. of  Eq.
(\ref{levy2}) it is presented by means of the Fourier integral
with a stretched exponential kernel. When
$F[x(\tau)]=\exp\{-\int_0^tV[x(\tau)]d\tau\}$, with $V(x)$ being a
potential, expression (\ref{levy2}) is a generalized Feynman-Kac
formula \cite{laskin1,ZT}

One should recognize that the general form of the path integral in
Eq. (\ref{levy2}) does not correctly describe systems with a
so-called topological constraint \cite{chaichian}. In this case
integrations over coordinates $x_1,x_2,\dots,x_N$ have finite
limits. As an example of such system, we consider fractional
quantum mechanics in an infinite well potential. We show that the
condition of the restriction of the integration in both the
formulation of the problem in Eq. (\ref{levy2}) and
correspondingly, in the fractional Riesz derivative must be taken
into account. Otherwise, the FSE for the infinite well potential
cannot be obtained.

\section{L\'evy quantum mechanics in potential
well}\label{sec:levy}

In complete analogy with the FFPE, fractional quantum mechanics
can be constructed from the  Feynman-Kac formula and $V(x)$ is the
potential. Following \cite{laskin1}, we determine a wave function
at the moment $t+\Delta t$ by means of Eq. (\ref{levy2})
\be\label{levy3} %
\psi(x,t+\Delta t)=\int_{-\infty}^{\infty}\frac{dp}{2\pi\hbar}\,
e^{-iD_{\alpha}\Delta t|p|^{\alpha}}
\int_{-\infty}^{\infty}dy\,e^{-\frac{i}{\hbar}p(x-y)}
e^{-\frac{i}{\hbar}V(y)\Delta t} \psi(y,t)\, .
\ee %
In the limit $\Delta t\rightarrow 0$, one obtains
\be\label{levy4} %
i\hbar\prt_t\psi(x,t)=\frac{D_{\alpha}}{2\pi\hbar}\int_{-\infty}^{\infty}dy
\int_{-\infty}^{\infty}
|p|^{\alpha}e^{-\frac{i}{\hbar}p(x-y)}\psi(y,t)dp +V(x)\psi(x,t)\,
.
\ee %
If one introduces the fractional Laplacian $(-\Delta)^{\alpha/2}$
by means of the Riesz fractional derivative \cite{SKM}
\be\label{riesz} %
(-i\hbar\prt_x)^{\alpha}\psi\equiv\hbar^{\alpha}(-\Delta)^{\alpha/2}\psi
=\frac{1}{2\pi\hbar}
\int_{-\infty}^{\infty}e^{-ipx}|p|^{\alpha}\phi(p)dp\, ,
\ee %
where $\phi(p)=\int_{-\infty}^{\infty}e^{ipy}\psi(y)dy$ is the
Fourier image, the fractional Schr\"odinger equation (FSE) is
obtained from Eq. (\ref{levy4})
\be\label{fse1} %
i\hbar\prt_t\psi=D_{\alpha}(-i\hbar)^{\alpha}
\prt_x^{\alpha}\psi+V(x)\psi\, .
\ee %
One sees that for $\alpha=2$, FSE (\ref{fse1}) reduces to a
conventional Hamiltonian mechanics with $D_2=1/2m$ being a half
inverse mass of a particle.

It should be noted that this equation is valid only for the smooth
potentials $V(x)$. In particular, for the infinite potential well
\be\label{well} %
V(x)=\left\{
\begin{array}{l}
0 \\
\infty
\end{array}
\begin{array}{l}
\mbox{if $\quad |x|\leq L$ } \\
\mbox{if $\quad |x|> L$}\, .
\end{array}
\right. \ee %
one cannot obtain the Schr\"odinger equation (\ref{fse1}) from
Eqs. (\ref{levy3}) and (\ref{levy4}). A simple explanation of this
problem relates to the absence of the expansion over $V\Delta t$
in Eq. (\ref{levy3}), since $\exp\left[-\frac{i}{\hbar}\Delta t
V(y)\right]$ is an infinitely fast oscillating function for any
arbitrary small but finite $\Delta t$. Note that one first
performs this expansion and then takes the limit $\Delta t=0$. To
overcome this obstacle, the cutoff of the limits of integration
over $y$ is performed. Moreover, this is a correct formulation of
the problem with the topological constraints, as the infinite
potential well is. This notion is also relevant for conventional
local quantum mechanics, see \textit{e.g.}, \cite{chaichian}.
Therefore, FSEs (\ref{levy4}) and (\ref{fse1}) are reduced to the
well defined problem of a free particle in a finite-size range
with the FSE
\be\label{fse2} %
i\hbar\prt_t\psi=\frac{D_{\alpha}}{2\pi\hbar}\int_{-L}^{L}dy
\int_{-\infty}^{\infty}
|p|^{\alpha}e^{-\frac{i}{\hbar}p(x-y)}\psi(y,t)dp \, ,
\ee %
which is furnished with the boundary conditions $\psi(x=\pm L)=0$.
Obviously, the Riesz fractional derivative (\ref{riesz}) can no
longer be used for the potential well, and in addition the Fourier
image $\phi(p)$ is not appropriately defined.

Let us perform the Fourier inversion in Eq. (\ref{fse2}) over
$k=p/\hbar$. This yields
\be\label{fse3} %
\frac{(\hbar)^{\alpha}}{2\pi}
\int_{-\infty}^{\infty}|k|^{\alpha}e^{-ik(x-y)}dk=
\frac{(-i\hbar)^{\alpha}}{2\pi}(-\prt_x)^2
\int_{-\infty}^{\infty}|k|^{\alpha-2}e^{-ik(x-y)}dk\, .
\ee %
One obtains this expression with the double differentiation for
$1<\alpha< 2$, while
for $0<\alpha<1$ one differentiates only once.
Therefore, we have the integration
\be\label{fse4} %
\int_{-\infty}^{\infty}|k|^{-\nu}e^{-ikz}dk=
2\int_0^{\infty}|k|^{-\nu}\cos kz=
\frac{2\pi|z|^{\nu-1}}{2\Gamma(\nu)\cos(\nu\pi/2)}\, ,
\ee %
where $\nu=2-\alpha$ for $1<\alpha\le 2$ and $\nu=1-\alpha$ for
$0<\alpha\le 1$, and $\Gamma(\nu)=(\nu-1)!$ is a gamma function.
To be specific, we consider $1<\alpha\le 2$ in the following
analysis, and the integration in Eqs. (\ref{fse3}) and
(\ref{fse4}) yields the fractional Laplace operator $\hat{\cal
L}^{\alpha}$ for the FSE (\ref{fse2}) in the form of the
Riemann-Liouville fractional derivative \cite{SKM}
\be\label{fse5} %
\hat{\cal L}^{\alpha}\psi(x)=
\frac{(\hbar)^{\alpha}D_{\alpha}}{2\Gamma(2-\alpha)\cos\frac{\alpha\pi}{2}}\prt_x^2
\int_{-L}^{L}|x-y|^{1-\alpha}\psi(y)\, dy\, .
\ee %
Note that
$$\frac{1}{2\Gamma(\nu)\cos\frac{\nu\pi}{2}}\int_a^b\frac{\phi(z)dz}{|x-z|^{1-\nu}}$$
is the Riesz fractional integral on the finite interval $[a,b]$
\cite{SKM} with $a\le x\le b$ and $0<\nu<1$. It can be presented
as the sum of the left and right Riemann-Liouville fractional
integrals
$$\int_a^x\frac{\phi(z)dz}{(x-z)^{1-\nu}}+\int_x^b\frac{\phi(z)dz}{(z-x)^{1-\nu}}\,
.$$

\subsection{Eigenvalue problem for the fractional Laplace
operator}\label{sec:eig}

Let us consider the eigenvalue problem
\be\label{eig1} %
\hL^{\alpha}\Psi_E=E\Psi_E
\ee %
with boundary conditions $\Psi_E(x=\pm L)=0$ that yields the
solution of FSE (\ref{fse2}) $\psi(x,t)=\sum_Ea_Ee^{-iEt}\Psi_E$,
where coefficients $a_E$ are defined by the initial conditions
$\psi_0(x)=\psi(x,t=0)$. We rewrite the fractional Laplace
operator in the form FSE (\ref{fse2}), which is convenient in the
following analysis,
\be\label{eig2} %
\hL ^{\alpha}f(y)=
D_{\alpha}\hbar^{\alpha}(i\prt_x)\int_{-L}^{L}\left[\frac{1}{2\pi}\int_{-\infty}^{\infty}
|k|^{\alpha-2}ke^{-ik(x-y)}dk\right]f(y)dy\, .
\ee%
Here we use again that on the real axis $|k|^2=k^2$. One easily
finds that the antisymmetric (odd) eigenfunction $\Psi_{E}^{\rm
odd}$ can be found in the form (see Appendix)
\be\label{eig3}  %
\Psi_E^{\rm odd}(x)=\Psi_m^{\rm
odd}(x)=\frac{1}{\sqrt{L}}\sin\frac{m\pi x}{L}\, , ~~~~m=1\, , 2\,
, \dots\, ,
\ee %
which satisfies the boundary condition $\Psi_m^{\rm odd}(x=\pm
L)=0$. Substituting solution (\ref{eig3}) with $f(y)=\Psi_E(y)$ in
Eq. (\ref{eig2}) and performing integration straightforwardly, one
obtains
$$\hL^{\alpha}\sin\frac{m\pi x}{L} =D_{\alpha} \left(\frac{\hbar
m\pi}{L}\right)^{\alpha}\sin\frac{m\pi x}{L}\, .$$ Therefore,
$\Psi_m^{\rm odd}(x)$ is the eigenfunction with corresponding
eigenvalue
\be\label{eig4} %
E^{\rm odd}\equiv E_m^{\rm odd}=D_{\alpha}\left(\frac{\hbar
m\pi}{L}\right)^{\alpha}\, .   %
\ee %
The same procedure is performed to find symmetric (even)
eigenfunctions
\be\label{eig5} %
\Psi_E^{\rm even}=\Psi_{2m+1}^{\rm even}
=\frac{1}{\sqrt{L}}\cos\frac{(2m+1)\pi}{2L}x
\ee %
with the eigenvalue
\be\label{eig6} %
E_m^{\rm even}=D_{\alpha}\left[\frac{\hbar
(2m+1)\pi}{2L}\right]^{\alpha}\, . \ee  %

\subsubsection{A comment on the ground state}

This result coincides with the solutions obtained in Ref.
\cite{laskin1}, but the crucial difference here is that solutions
(\ref{eig3}) - (\ref{eig6}) belong to a completely different
fractional operator (\ref{fse5}), which is defined on the
finite-size range of the potential well. Note also that in this
case a deficiency with the ground state
$\frac{1}{\sqrt{L}}\cos\frac{\pi x}{2L}$, correctly stated in
Refs. \cite{Jeng,Hawkins}, no longer exists, since it is obtained
by the appropriately determined operator.

\section{Fractional path integral: fractional quantum mechanics
with topological constraints}\label{sec:fpi}

One should bear in mind that the Feynman-Kac formula in Eq.
(\ref{levy3}) and FSE (\ref{levy4}) are introduced in unbounded
Euclidian (configuration) space. In contrast, the configuration
space of a particle in the infinite potential well is contracted
to the finite size of the  potential well. This leads to the
quantum mechanics in a space with topological constraints. Even
for the conventional local quantum mechanics, the path integral
presentation is not an easy task, as stated in \cite{chaichian}.
Fortunately, since eigenvalue solutions (\ref{eig1})-(\ref{eig5})
are known, the path integral for the L\'evy process in the finite
area can be constructed. Therefore, one can obtain Eq.
(\ref{fse1}) with infinite limits of integration for the Riesz
operator. Let us define for convenience $|x\rgl \equiv\Psi_m^{\rm
odd}(x)$. Then, Eq. (\ref{levy3}) can be rewritten:
\be\label{fpi1} %
\psi(x,\Delta t)\equiv\lgl x|\psi(\Delta t)\rgl=\lgl
x|e^{-i\hL^{\alpha}\Delta t/\hbar}|\psi_0\rgl  % = \nonumber \\
 = \int_{-L}^{L}dx_1 \lgl x|e^{-i\hL^{\alpha}\Delta
t/\hbar}|x_1\rgl\lgl x_1|\psi_0\rgl\, . \ee  %
We focus on the evolution Green's function
\be\label{fpi2}  %
G(x,\Delta t;x_1)=\lgl x|e^{-i\hL^{\alpha}\Delta t/\hbar}|x_1\rgl
=\int_{-\infty}^{\infty}\frac{dk}{2\pi}e^{-i|k|^{\alpha}D_{\alpha}\Delta
t}\int_{-L}^{L}dye^{-ik(x_1-y)}\lgl x|y\rgl\, . \ee  %
Taking into account that
\be\label{fpi3} %
\lgl x|y\rgl=\frac{1}{L}\sum_{n=0}^{\infty}\sin(k_nx)\sin(k_ny)\,
~~~~~~k_n\equiv\frac{\pi n}{L} \ee  %
and the Poisson summation formula
\be\label{fpi4} %
\sum_{l=-\infty}^{\infty}e^{2\pi i
xl}=\sum_{m=-\infty}^{\infty}\delta(x-m)\, ,  \ee  %
one obtains (see also \cite{chaichian})
\bea\label{fpi5}  %
G(x,\Delta t;x_1)&=&\frac{1}{2}\sum_{l=-\infty}^{\infty}
\int_{-\infty}^{\infty}\frac{dk}{2\pi}e^{-i\hbar^{\alpha-1}|k|^{\alpha}D_{\alpha}\Delta
t}\left[e^{ik(x-x_1+2Ll)}-e^{ik(x+x_1+2Ll)}\right] \nonumber \\
&=& \frac{1}{2}\sum_{z=\pm x}\sum_{l=-\infty}^{\infty}
\int_{-\infty}^{\infty}\frac{dk}{2\pi}
 e^{-i\hbar^{\alpha-1}|k|^{\alpha}D_{\alpha}\Delta t}  \nonumber \\
&\times&\exp\left\{ik(z-x_1+2Ll)+i\pi[\theta(-z)-\theta(x_1)]\right\}\, , \eea %
where the $\theta$ functions in the exponential provide the
correct signs for $z=\pm x$. Now, we take into account
\be\label{fpi6}  %
\frac{1}{2}\sum_{z=\pm x} \sum_{l=-\infty}^{\infty}
\int_{-L}^{L}dx_1 \longleftrightarrow
\int_{-\infty}^{\infty}dx_1\, , \ee %
which returns one to the fractional integration with infinite
limits, as in Eq. (\ref{levy3}). The essential difference here is
the appearance of the topological phase due the $\theta$ functions
in Green's function (\ref{fpi5}). Continuing this procedure by
repeating it $N$ times ($t=N\Delta t$), one arrives at the correct
analogy of Eq. (\ref{levy2}) for the infinite well potential
\bea\label{fpi7}  %
&G(x,t;x_0)=\sum_{l=-\infty}^{\infty}\sum_{z=\pm x+2lL}
\prod_{j=1}^N\int_{-\infty}^{\infty}dx_j
\prod_{j=1}^{N+1}\int_{-\infty}^{\infty}\frac{dk_j}{2\pi}
\nonumber \\
&\times
\exp\{i\sum_{j=1}^{N+1}\left[k_j(x_j-x_{j-1})+\pi\theta(-x_j)-\pi\theta(-x_{j-1})-
\hbar^{\alpha-1}|k_j|^{\alpha}D_{\alpha}\Delta t\right]\}\, ,  \eea %
where $x_{N+1}\equiv z$. The topological term is a purely boundary
expression\footnote{For $\alpha=2$, this expression coincides with
the results in Ref. \cite{chaichian}. For the self-contained
presentation, we shall adjust some formulae thereof. Namely,
taking the continuous time limit, one arrives at the path integral
$ G(x,t;x_0)=\frac{1}{2}\sum_{l=-\infty}^{\infty}\sum_{z=\pm
x+2lL} \int Dx(\tau)\int \frac{D p(\tau)}{2\pi\hbar}
\exp\left\{\frac{i}{\hbar}S\right\}$, where the action
$S[z]=\int_0^td\tau[p\dot{x}-H(|p|)-\hbar\pi\dot{x}\delta(x)]$
contains the new topological term $ S_{\rm
top}[z]=-\hbar\pi\int_0^td\tau[\dot{x}\delta(x)]=
\hbar\pi[\theta(-z)-\theta(-x_0)]$, which is a purely boundary
expression.}, and therefore, integration over all $x_j$ in Eq.
(\ref{fpi7}) yields a product of $\delta$ functions
$\prod_{j=1}^N\delta(k_j-k_{j+1})$, which finally yields the
Fourier inversion of the Green function in the form of Fox's
$H_{2,2}^{1,1}$ function \cite{laskin1,WGMN} in its Fourier form
\be\label{fpi8}  %
G(x,t;x_0)=\sum_{l=-\infty}^{\infty}\frac{1}{4\pi}\int_{-\infty}^{\infty}dk
e^{-i\hbar^{\alpha-1}|k|^{\alpha}D_{\alpha}
t}\left[e^{ik(x-x_0+2lL)}-e^{-ik(x+x_0+2lL)}\right]
%\nonumber \\
%\sum_{l=-\ifty}^{\infty}\frac{1}{\alpha|x_l^-|}
%H_{2,2}^{1,1}\left\{\frac{1}{\hbar}\left(\frac{\hbar}{D_{\alpha}t}\right)^{\frac{1}{\alpha}
%|x_l^-|\large|
\, . \ee %
Because of the properties of $H_{2,2}^{1,1}$ Fox's function, at
$\alpha=2$, the Green function (\ref{fpi8}) reduces to the free
particle in the box with the Green function \cite{chaichian}
\be\label{fpi9} %
{\lgl x,t|x_0,0\rgl}_{\rm box}=\sum_{l=-\infty}^{\infty}
\frac{1}{\sqrt{8i\pi\hbar t}}
\left[e^{\frac{im}{2\hbar}\frac{(x-x_0+2lL)^2}{t}}-
e^{\frac{-im}{2\hbar}\frac{(x+x_0+2lL)^2}{t}}
\right]\, . \ee  %

\section{Example: Adiabatic approximation in moving
walls}\label{sec:exm}

Handling the path integral expression in such a simple form as Eq.
(\ref{fpi8}), one can consider a system of a L\'evy particle
inside moving walls. For $\alpha=2$, this problem corresponds to
the so-called Fermi acceleration \cite{chirzas}, where chaotic
dynamics can be realized because of the interaction of nonlinear
resonances. However, for $1<\alpha<2$,  the physical
implementation of the classical limit of the problem is vague,
since a particle is not free. Therefore, the interaction of
nonlinear resonances is not considered. Here, the problem is
treated in the adiabatic approximation. In this case, Eq.
(\ref{fpi8}) can be used for further quantum mechanical analysis.
Let the infinite walls at $x=\pm L$ move periodically, such that
the boundary conditions for the wave function are $\psi(x=\pm
L(t))=0$, where $L(t)=L+\vep\sin(\nu t)$ with $\vep/L\ll 1$. To
calculate the density of states, we are interested in the trace of
the Green function
\be\label{exm1} %
g(t)=\int_{-L}^{L}dx_0G(x_0,t;x_0)  %\equiv \int dx_0G_0(t)\, ,
\ee %
where $ G(x=x_0,t;x_0)$ can be obtained from Eq. (\ref{fpi8}) in
the form
\be\label{exm2} %
G(x_0,t;x_0)=
\sum_{l=-\infty}^{\infty}\frac{1}{4\pi}\int_{-\infty}^{\infty}dk
e^{-i\hbar^{\alpha-1}|k|^{\alpha}D_{\alpha} t}
\left[e^{ik(2lL)}-e^{-ik(2x_0+2lL)}\right] \, .
\ee %
Performing summation over $l$ by the Poisson summation formula
\be\label{exm3} %
\sum_{l=-\infty}^{\infty}e^{\pm
2iklL}=1+2\sum_{l=1}^{\infty}e^{2iklL}=\frac{\pi}{L}
\sum_{m=-\infty}^{\infty}\delta\left(k-\frac{\pi m}{L}\right)\, , %
\ee %
one can perform integration over $k$, which leads to summation
over the spectrum $k=k_m=\pi m/L$. Then, performing integration
over $x_0$, one obtains for the trace
\be\label{exm4} %
g_0(t)=\sum_{m=1}^{\infty}e^{-i|\hbar\pi
m/L|^{\alpha}D_{\alpha}t/\hbar} \, .
\ee %
One also obtains this result from the expression for the Green
function $G(x_0,t;x_0)=\sum_n \exp(-iE_n^{\rm
odd}t/\hbar)\Psi_n^{\rm odd}(x_0)\Psi_n^{*\, \rm odd}(x_0)$. The
Fourier transform $\tilde{g}_0(E)=\hF[g_0(t)]$ yields the density
of states $\rho_0(E)=-\frac{1}{\pi}\Im\tilde{g}(E)$
\be\label{exm5} %
\rho_0(E)=\hbar\sum_m\delta\left(E-D_{\alpha}|\hbar
k_m|^{\alpha}\right)\,
, \ee  %
where $E\equiv E^{\rm odd}$, which satisfies the eigenvalue
problem, discussed in Sec.~\ref{sec:eig}.

In the case of the moving walls, at the replacement $L\rightarrow
L(t)$, the adiabatic approximation makes it possible to obtain the
trace of the Green's function in the form
\be\label{exm6} %
g_{\vep}(t)=e^{(\vep\sin\nu t) \frac{d}{dL}}g_0(t)=
\sum_me^{-i(\pi m/L(t))^{\alpha}D_{\alpha}t}\, , \ee %
where we take $\hbar\equiv 1$. Let us present a shift operator in
the form
\be\label{exm7} %
e^{(\vep\sin\nu t)\frac{d}{dL}}=\sum_{n=-\infty}^{\infty}
\frac{1}{2\pi}\int_{-\pi}^{\pi}d\xi e^{\vep\sin\xi\frac{d}{dL}}
e^{in(\xi-\nu t)}\, .
\ee %
Then, the Fourier transform over time can be easily performed,
which yields the following density of states
\be\label{exm8}  %
\rho_{\vep}(E)=\sum_{n=-\infty}^{\infty}
\frac{1}{2\pi}\int_{-\pi}^{\pi}d\xi e^{(\vep\sin\xi)\frac{d}{dL}}
e^{in\xi} 2L\sum_m\delta\left(E-n\nu-D_{\alpha}\left|\frac{\pi
m}{L}\right|^{\alpha}\right)\, .  \ee %
%Obviously, integration over $\xi$ destroys the structure of the summation over $n$.
Parameter $\xi$ determines the quasi-energy spectrum, namely, its
band. Therefore, for the moving walls the discrete spectrum
(\ref{eig4}) changes essentially:
\be\label{exm9} %
E_m\rightarrow E_{m,n}(\xi)=n\nu+D_{\alpha}\left|\frac{\pi
m}{L+\vep\sin\xi}\right|^{\alpha}\approx
n\nu+D_{\alpha}\left|\frac{\pi m}{L}\right|^{\alpha}-\vep\alpha
D_{\alpha}\frac{|\pi m|^{\alpha}}{L^{\alpha+1}}\sin\xi\, . \ee  %
This solution corresponds to the reconstruction of the energy
spectrum of the stationary problem (\ref{eig1}) to the quasienergy
spectrum with a narrow band structure.

\section{Conclusion}

The introduction of the quantum L\'evy process in a box leads to
the need to account for the topological constraints in the space.
This problem can be treated in the framework of the path integral
formalism with the L\'evy measure. A correct path integral
consideration is possible when the eigenvalue problem (\ref{eig1})
is appropriately defined for the infinite well potential
(\ref{well}), and the eigenfunctions are known, or can be obtained
for the fractional (nonlocal) operator (\ref{fse5}). An analytical
expression for the evolution operator is obtained in the path
integral presentation. For $\alpha=2$, the path integral has the
limit of the correct expression, which is well known in local
quantum mechanics with topological constraints \cite{chaichian}.
Although the results of the eigenvalue problem coincide with the
solutions obtained in Ref. \cite{laskin1}, the solutions
(\ref{eig3}) - (\ref{eig6}) belong to the fractional operator
(\ref{fse5}), which differ from those obtained in \cite{laskin1},
since these are defined on the finite-size range of the potential
well and correspond to the L\'evy walks. In this case, the ground
state $\frac{1}{\sqrt{L}}\cos\frac{\pi x}{2L}$ is correctly
defined as well, and to some extent, this ends the discussion on
the ground states (see Refs. \cite{Jeng,Hawkins}).

An important point of any ``fractional quantum mechanics'' is the
classical limit $\hbar\rightarrow 0$, where a serious question
emerges about the physical meaning of its classical counterpart.
Obviously, a quantum L\'evy particle is not a free particle in the
classical limit. As an example of such a possible realization, the
L\'evy process is considered for oscillating walls. For
$\alpha=2$, this problem corresponds to the so-called Fermi
acceleration \cite{chirzas} leading to classical and quantum
chaos. But for $1<\alpha<2$,  the physical implementation of the
classical limit of the problem is vague, since a particle is not
free. Therefore, the interaction of nonlinear resonances is not
considered here. The problem is treated in the adiabatic
perturbative approach. This question can be an interesting task
for future studies of classical and quantum chaos of the Fermi
acceleration mechanism in the nonlocal quantum mechanics.

This research was supported by the Israel Science Foundation
(ISF).

\appendix
\section{Eigenvalue problem}\label{sec:app_A}

\def\theequation{A. \arabic{equation}}
\setcounter{equation}{0}

Let us consider the eigenvalue problem (\ref{eig1}) for the
fractional Laplace operator and find the antisymmetric (odd)
eigenfunctions $\Psi_{E}^{\rm odd}$ in the form
\be\label{App_1}  %
\Psi_E^{\rm odd}(x)=\Psi_m^{\rm odd}(x)
=\frac{1}{\sqrt{L}}\sin\frac{m\pi x}{L}\, , ~~~~m=1\, , 2\, ,
\dots\, ,
\ee %
which satisfies the boundary condition $\Psi_m^{\rm odd}(x=\pm
L)=0$.
We rewrite the fractional Laplace operator in the form FSE
(\ref{fse2}), which is convenient in the analysis,
\be\label{App_2} %
\hL ^{\alpha}\Psi_m^{\rm odd}(x)=
D_{\alpha}\hbar^{\alpha}(i\prt_x)\int_{-L}^{L}\left[\frac{1}{2\pi}\int_{-\infty}^{\infty}
|k|^{\alpha-2}ke^{-ik(x-y)}dk\right]\Psi_m^{\rm odd}(y)dy\, .
\ee%
Let us first perform integration over $y$. This yields
\be\label{App_3}  %
\frac{1}{2i}\int_{-L}^{L}\left[e^{iky+iz}-e^{iky-iz}\right]
=(-1)^m\frac{2\pi m}{iL}\frac{\sin(kL)}{(k+z)(k-z)}\, ,
\ee %
where $z=\pi my/L$. The next step is integration over $k$.  From
Eqs. (\ref{App_2}) and (\ref{App_3}), we have integrals
\be\label{App_4}  %
\frac{1}{2\pi} \int_{-\infty}^{\infty}
\frac{k|k|^{\alpha-2}}{(k+z)(k-z)}\left[e^{ik(L-x)}-e^{-i(L+x)}\right]
\equiv I^{(+)}-I^{(-)}\, ,  \ee  %
where sign $(+)$ corresponds to the analytical continuation in the
upper half plain, while $(-)$ corresponds to the analytical
continuation in the lower half plain. Using the Residue theorem,
one obtains that the integration yields
\be\label{App_5} %
I^{(+)}-I^{(-)}=i(-1)^m\left(\frac{m\pi}{L}\right)^{\alpha-2}
\cos\left(\frac{m\pi}{L}x\right)\, . \ee  %
Acting on this result by the rest part of the operator, which
reads
$(-1)^mD_{\alpha}\hbar^{\alpha}\left(\frac{m\pi}{L}\right)(i\prt_x)$,
one obtains
\be\label{App_6} %
\hL ^{\alpha}\Psi_m^{\rm odd}(x)=
D_{\alpha}\hbar^{\alpha}\left(\frac{m\pi}{L}\right)^{\alpha}
\Psi_m^{\rm odd}(x)\equiv E_m^{\rm odd}\Psi_m^{\rm odd}(x)\, . \ee  %


\begin{thebibliography}{00}

\bibitem{laskin1} N. Laskin, Fractals and Quantum Mechanics,
Chaos  10: 780, 2000.

\bibitem{Jeng} M. Jeng, S.-L.-Y. Xu, E. Hawkins, and J. M. Schwarz,
On the nonlocality of fractional Schr\"odinger equation, J. Math.
Phys. 51: 062102, 2010.

\bibitem{Bayin1} S.S. Bayin, On the consistency of the solutions of the space
fractional Schr\"odinger equation, J. Math. Phys. 53: 042105,
2012.

\bibitem{Bayin2} S.S. Bayin, Comment on ``On the consistency of the
solutions of the space fractional Schr\"odinger equation'', J.
Math. Phys. 53: 084101, 2012.

\bibitem{Hawkins} E. Hawkins and J.M. Schwarz, Comment on
``On the consistency of solutions of the space fractional
Schr\"odinger equation'' [J. Math. Phys. 53 042105, 2012], J.
Math. Phys. 54: 014101, 2013.

\bibitem{luchko} Yu. Luchko, Fractional Schr\"odinger equation for a particle moving
in a potential well, J. Math. Phys. 54: 012111, 2013.

\bibitem{herrmann} R. Herrmann, The fractional Schr\"odinger equation
and the infinite potential well - numerical results using the
Riesz derivative, Gam. Ori. Chron. Phys. 1: 1,  2013.

\bibitem{Dubiec} B. Dybiec, Gudowska-Nowak, and P. H\"anggi,
L\'evy-Brownian motion on finite intervals: Mean first passage
time analysis, Phys. Rev. E 73: 046104 2006.

\bibitem{Barthelemy} P. Barthelemy, J. Bertolotti, D. Wiersma, A L\'evy flight for
light, Nature 453: 495, 2008.

\bibitem{west} B.J. West, Quantum L\'evy Propagators, J. Phys. Chem. B
104: 3830, 2000.

\bibitem{feynman} R.P. Feynman and A.R. Hibbs, \textit{Quantum Mechanics
and Path Integrals}. McGraw--Hill, New York, 1965.

\bibitem{kac} M. Kac, \textit{Probability and Related Topics in Physical
Sciences}. Interscience, New York, 1959.

\bibitem{shlesinger} E.W. Montroll and M.F. Shlesinger,
The wonderful world of random walks, in J. Lebowitz and E.W.
Montroll  (eds) \textit{Studies in Statistical Mechanics}, v. 11
Noth--Holland, Amsterdam, 1984.

\bibitem{klafter} R. Metzler and J. Klafter,
The random walk guide to anomalous diffusion: A fractional
dynamics approach, Phys. Rep. 339: 1, 2000.

\bibitem{ZT} V.E. Tarasov and G.M. Zaslavsky, Fractional generalization of Kac
integral, Comm. Nonlinear Scien.  Num. Simul. 13: 248, 2008.

\bibitem{chaichian} M. Chaichian and A. Demichev, \textit{Path
Integrals in Physics}, Institute of Physics Publishing, Bristol,
2001, v.1.

\bibitem{SKM}  S.G. Samko, A.A. Kilbas, and O.I. Marichev,
\textit{Fractional Integrals and Derivatives: Theory and
Applications}, Gordon and Breach, New York, 1993.

\bibitem{WGMN} B.J. West, P. Grigolini, R. Metzler, and T.F.
Nonnenmacher, Fractional diffusion and L\'evy stable processes,
Phys. Rev. E 55: 99, 1997.

\bibitem{chirzas} B.V. Chirikov and G.M. Zaslavsky,
On the mechanism of one-dimensional Fermi acceleration,
\textit{Dokl. Akad. Nauk SSSR} 159: 306 1964.



\end{thebibliography}
\end{document}